\documentstyle[pra,aps]{revtex}
\begin{document}
\draft
\title{Stark-Induced Electric Dipole Amplitudes for Hyperfine Transitions}
\author{David DeMille}
\address{Department of Physics, Amherst College, Amherst, MA 01002}
\author{M. G. Kozlov}
\address{Petersburg Nuclear Physics Institute,\\
Gatchina, Leningrad District, 188350, Russia}
\date{\today}
\maketitle

\begin{abstract}
Stark-induced electric dipole amplitudes between states of the same nominal
parity can be important in experiments to observe parity nonconservation in
atoms. The Stark-induced E1 amplitudes are expressed in terms of an
irreducible spherical-tensor decomposition. This formalism is applied to the
specific case of transitions between hyperfine sublevels of a single atomic
state. It is shown that in the ground states of alkali atoms, such
transitions are suppressed by many orders of magnitude relative to naive
expectations.
\end{abstract}

\pacs{32.60.+i, 32.70.Cs, 32.30.Bv}


\section{Introduction}


The first definitive observation of a nuclear spin-dependent (NSD)
contribution to atomic parity nonconservation (PNC) was recently reported
\cite{Wood}. This effect arises primarily because of an electromagnetic
interaction between atomic electrons and the nucleus. In particular, because
of PNC interactions {\em within} the nucleus, the nucleus acquires a P-odd
electromagnetic multipole moment known as the anapole moment \cite
{Zeldovich,NuclearAnapole}. Measurements of atomic NSD-PNC amplitudes can
thus be interpreted in terms of the nuclear anapole moment. The nuclear
anapole moment itself can be related to the strengths of various hadronic
PNC couplings. These couplings are in general poorly determined, and the
available data are in some cases contradictory. Thus, there is considerable
interest in additional measurements of atomic NSD-PNC, as a means to probe
hadronic neutral-weak interactions \cite
{Flambaumanapolereview,HaxtonAnapoleReview}.

NSD-PNC effects are extremely small, and their observation requires both
extreme statistical sensitivity and careful control over minute systematic
effects. Gorshkov {\it et al}$.$ have proposed a novel technique for
observing NSD-PNC\cite{Gorshkov}. They suggest using a strong microwave
electric field to drive PNC-induced electric dipole (E1) transitions between
hyperfine sublevels of a single atomic state. Interference between this E1
transition amplitude and a (parity-allowed) M1 amplitude driven by a weak
microwave magnetic field (coherent with the electric field) gives rise to a
PNC observable. This technique offers the promise for unprecedented
statistical sensitivity. However, Gorshkov {\it et al}$.$ give only a
limited and qualitative analysis of possible systematic effects in the
measurement they propose.

Further consideration of the proposal of Ref. \cite{Gorshkov} has led us to
investigate the phenomenon of Stark-induced (SI) E1 transitions between
hyperfine sublevels. These arise in the presence of a DC electric field,
which mixes states of opposite parity into the initial and final states;
thus E1 transitions can be induced between one nominal state and the admixed
component of the other state\cite{BouchiatPNCProposal2}. Such transitions
can be of importance in PNC measurements for two reasons. First, the
presence of stray electric fields leads to SI-E1 transition amplitudes
between hyperfine sublevels, which can mimic the effect of the PNC-induced
E1 amplitudes; the size of the SI amplitudes relative to the PNC-induced
amplitudes determines the extent to which stray fields must be controlled.
Second, by deliberately applying a large electric field, the SI-E1 amplitude
could in principle be used as the primary parity-allowed amplitude, against
which the PNC amplitude can interfere\cite{BouchiatPNCProposal2}. This
technique gives excellent control over a variety of systematic effects, and
indeed is the principle of the measurement of Ref. \cite{Wood}, in which the
optical 6s-7s transition in Cs was studied.

We show here that in the particular case of alkali atoms (as discussed in
Ref. \cite{Gorshkov}), the SI-E1 amplitudes between hyperfine sublevels of
the ground state are suppressed by many orders of magnitude relative to
naive expectations. This conclusion means that it is impossible in practice
to use the SI amplitudes as the primary parity-allowed amplitude in
experiments of the type proposed by Gorshkov {\it et al.} On the other hand,
it also means that systematic effects due to stray electric fields in such
experiments are negligibly small under reasonable conditions.


\section{General argument for the suppression of SI-E1 hyperfine transitions}


The SI-E1 transition $a\rightarrow b$ in a constant field ${\bf E}$ and
time-varying field ${\bf \bbox{\varepsilon}}$ has amplitude
\begin{equation}
W_{ba}=\sum_{q,q^{\prime }}(-1)^{q+q^{\prime }}A_{q,q^{\prime
}}E_{-q}\varepsilon _{-q^{\prime }},  \label{a1}
\end{equation}
where
\begin{equation}
A_{q,q^{\prime }}=e^{2}\sum_{n}\left( \frac{\langle b|r_{q}|n\rangle \langle
n|r_{q^{\prime }}|a\rangle }{E_{b}-E_{n}}+\frac{\langle b|r_{q^{\prime
}}|n\rangle \langle n|r_{q}|a\rangle }{E_{a}-E_{n}}\right) ,  \label{a2}
\end{equation}
and the subscripts $q,q^{\prime }$ refer to spherical vector components.

The SI-E1 transition can be usefully understood as a special case of a
two-photon transition, with one photon at zero frequency. For electric
dipole interactions, each photon has orbital angular momentum $L=0$ with
respect to the atom. Thus, since each photon has spin $S=1$, the two-photon
system must have total angular momentum $J_{{\rm tot}}=S_{{\rm tot}}=0,1,$
or 2. These three cases correspond to scalar ($\alpha _{\rm St}$), vector ($%
\beta _{\rm St}$), and tensor ($\gamma _{\rm St}$) transition amplitudes,
respectively. We are primarily interested here in transitions between
sublevels of the $s_{1/2}$ ground state of an alkali atom. For this case,
each of the three possible amplitudes is strongly suppressed (or vanishes)
relative to the naive estimate of the amplitude: $A\sim e^{2}a_{0}^{2}/{\cal %
R}\sim a_{0}^{3}$ (where $a_{0}$ is the Bohr radius and ${\cal R}$ is the
Rydberg constant).  These suppressions can be understood in terms of
selection rules for two-photon transitions, which were derived in Ref. \cite
{Bonin}. We begin by considering these (approximate) selection rules for the
specific case of hyperfine transitions, and estimate the residual values of
the SI-E1 amplitudes.

First, we note that the scalar amplitude $\alpha _{\rm St}$ vanishes for
hyperfine transitions. This is a trivial consequence of the fact that the
scalar amplitude cannot change any angular quantum numbers, and thus cannot
contribute to any transitions between sublevels of the same state. This
argument fails when the principal quantum number changes; thus, for the
well-studied 6s-7s transition in Cs, $\alpha _{\rm St}(6s-7s)=269$ $a_{0}^{3}$%
\cite{DzubaAlpha}. (This is somewhat larger than the naive estimate given
above because the radial matrix elements in Cs and other alkalis are $\sim
5a_{0}$, and the energy denominators are $\sim 0.1{\cal R}$.)

Next, we show that for hyperfine transitions, the vector amplitude $\beta
_{\rm St}$ is suppressed by a factor $\sim \frac{m_{e}}{m_{p}}Z\alpha ^{2}$,
where $\alpha $ is the fine-structure constant. This suppression can be
understood as a consequence of Bose statistics for photons. For S$_{{\rm tot}%
}$=1, the spin component of a two-photon wavefunction is antisymmetric under
interchange of the two photons. Thus, the space-time component of the
wavefunction must also be antisymmetric under interchange, in order to
satisfy Bose statistics. In the electric-dipole approximation, where the
spatial part of the wavefunction is constant, the total wavefunction must
therefore vanish if the photons have the same energy. (This statement is
closely related to the Landau-Yang theorem of high-energy physics, which
states that the decay of a J=1 particle into two photons is forbidden \cite
{Landau,Yang}; this result has also been derived for atomic two-photon
transitions \cite{Bonin}.) In the limit where the frequencies of the two
photons are close but not identical, the vector part of the two-photon
transition amplitude does not vanish, but is suppressed by a factor $\sim
\frac{\hbar \omega _{1}-\hbar \omega _{2}}{E_{v}-E_{r}}$, where $\hbar
\omega _{i}$ is the energy of the $i$th photon, and $E_{v}-E_{r}$ is the
difference in energy between a virtual intermediate state and a real
intermediate state \cite{BowersYbLifetimes}. For the case of SI-E1
transitions between hyperfine sublevels of the same state, $\hbar \omega
_{1}-\hbar \omega _{2}$ = $\hbar \omega $ = $E_{{\rm hfs}}$ (the hyperfine
splitting energy), $E_{v}$ = $E_{{\rm hfs}}$ (or 0), and $E_{r}$ = $E_{{\rm %
el}}$ (the energy splitting between electronic states). Thus, for the case
of interest here, $\beta _{\rm St}$ is suppressed by an additional factor $\sim
\frac{E_{{\rm hfs}}}{E_{{\rm el}}}\sim \frac{m_{e}}{m_{p}}Z\alpha ^{2}$ (see
e.g. \cite{Sobelman}). Clearly, this suppression is absent for the vector
component of optical SI-E1 transitions such as the 6s-7s transition in Cs.

For the particular case of transitions between hyperfine sublevels of the
ground state of an alkali atom, the vector amplitude is suppressed by an
additional factor of $\sim Z^{2}\alpha ^{2}$. In the absence of both
electron and nuclear spin, $\beta _{\rm St}$ must vanish for any
$s-s^{\prime }$
($0-0^{\prime }$) transition. However, the addition of electron spin alone
is sufficient to allow this amplitude, since then the transition is of the
type $\frac{1}{2}-\frac{1}{2}^{\prime }$. (Nuclear spin alone will also
induce a nonzero value of $\beta _{\rm St}$, but this is a much smaller
effect.)
Since for any $s_{1/2}-s_{1/2}^{\prime }$ transition $\beta _{\rm St}$
explicitly relies on the presence of electron spin, it must be that $\beta
_{\rm St}$ is suppressed relative to the naive estimate by
$\sim \frac{E_{{\rm fs%
}}}{E_{v}-E_{r}}\sim \frac{E_{{\rm fs}}}{E_{{\rm el}}}\sim Z^{2}\alpha ^{2}$%
, where $E_{{\rm fs}}$ is a fine-structure energy splitting (see e.g. \cite
{Sobelman}). This suppression is indeed present for the 6s-7s transition in
Cs\cite{BouchiatPNCProposal2}, where the ratio of vector to scalar
amplitudes is $\beta _{\rm St}/\alpha _{\rm St}=0.1$
\cite{ChoAlphaOverBeta}. For the
hyperfine transitions of interest here, the vector amplitude is thus
suppressed relative to its naive value by an overall factor $\sim \frac{m_{e}%
}{m_{p}}Z^{3}\alpha ^{4}.$

Finally, we consider the tensor part $\gamma _{\rm St}$ of the SI-E1 amplitude.
For $s-s^{\prime }$ transitions this amplitude vanishes even in the presence
of electron spin (i.e., for a $\frac{1}{2}-\frac{1}{2}^{\prime }$
transition); its presence relies explicitly on the presence of nuclear spin.
Thus, $\gamma _{\rm St}$ must be suppressed relative to the naive value by a
factor $\sim \frac{E_{{\rm hfs}}}{E_{v}-E_{r}}\sim \frac{E_{{\rm hfs}}}{E_{%
{\rm el}}}$ $\sim \frac{m_{e}}{m_{p}}Z\alpha ^{2}$. Let us make this
statement more explicit by noting that, for any nuclear spin $I\geq 1$,
nuclear spin alone is sufficient to induce a non-zero tensor amplitude. (For
all stable alkali-atom nuclei, $I\geq 1$.) Since electron spin is neither
necessary nor sufficient to produce a tensor amplitude, we expect that its
presence should not significantly alter the magnitude of $\gamma _{\rm St}$.
Rather, the tensor amplitude can be understood from the hypothetical case of
an atom with nuclear spin, but no electron spin. In this case, the only
relevant hyperfine splitting is that due to the interaction of $p$-states
with the nuclear magnetic dipole and electric quadrupole moments. Thus,
$\gamma _{\rm St}$ is suppressed relative to its naive value by a factor of
$\sim \frac{E_{{\rm hfs}}\left( p\right) }{E_{{\rm el}}}$.
This argument holds
even for optical $s-s^{\prime }$ transitions such as the 6s-7s transition in
Cs; for this reason, the amplitude $\gamma _{\rm St}$ is negligibly small for
this transition, and is always ignored. We note in passing that the
suppression of $\gamma _{\rm St}$ is not present at all for transitions
between hyperfine sublevels of any electronic state with $J\geq 1$.
However, we confine ourselves to the particular case of hyperfine
transitions in alkali-atom ground states.


\section{Expressions for scalar, vector and tensor polarizabilities}


We now discuss specific calculations of the SI-E1 amplitudes. For our
purposes it is convenient to rewrite the general second-rank tensors $%
A_{q,q^{\prime }}$ and $E_{-q}\varepsilon _{-q^{\prime }}$ of (\ref{a1}) and
(\ref{a2}) in terms of their irreducible tensor components $A_{Q}^{K}$ and $%
(E\otimes \varepsilon )_{-Q}^{K}$. We use the standard transformations\cite
{Sobelman}, e.g.
\begin{equation}
A_{Q}^{K}=(-1)^{Q}\sqrt{2K+1}\sum_{q,q^{\prime }}\left(
\begin{array}{ccc}
1 & 1 & K \\
-q & -q^{\prime } & Q
\end{array}
\right) A_{q,q^{\prime }},  \label{a7}
\end{equation}
and
\begin{equation}
E_{-q}\varepsilon _{-q^{\prime }}=\sum_{K,Q}(-1)^{Q}\sqrt{2K+1}\left(
\begin{array}{ccc}
1 & 1 & K \\
-q & -q^{\prime } & Q
\end{array}
\right) (E\otimes \varepsilon )_{-Q}^{K}  \label{a5}
\end{equation}
so that the transition probability (\ref{a1}) is written as the contraction
of irreducible tensor components (as was done for the general case of
two-photon transitions in Ref. \cite{Bonin}):
\begin{equation}
W_{ba}=\sum_{K,Q}(-1)^{Q}A_{Q}^{K}(E\otimes \varepsilon )_{-Q}^{K}.
\label{a6}
\end{equation}

We further rewrite the irreducible tensor component $A_{Q}^{K}$ in terms of $%
A^{K}$, the reduced amplitude of rank $K:$
\begin{equation}
A_{Q}^{K}=(-1)^{J_{b}-M_{b}}\left(
\begin{array}{ccc}
J_{b} & K & J_{a} \\
-M_{b} & Q & M_{a}
\end{array}
\right) A^{K}.  \label{a7a}
\end{equation}
In the general case the transition amplitude is fully described by three
independent reduced amplitudes with $K=0,1,2$.

The SI-E1 transition amplitude has traditionally been written in terms of
the quantities $\alpha _{\rm St}$, $\beta _{\rm St}$, and $\gamma _{\rm St}$,
such that

\widetext
\begin{equation}
W_{ba}=\alpha _{\rm St}{\bf E}\cdot {\bf \bbox{\varepsilon}}
+{\bf \bbox{\beta}}%
_{\rm St}\cdot {\bf E}\times {\bf \bbox{\varepsilon}}
+\gamma _{\rm St}^{i,k}(\frac{1%
}{2}E_{i}\varepsilon _{k}+\frac{1}{2}E_{k}\varepsilon _{i}
-\frac{1}{3}{\bf E}%
\cdot {\bf \bbox{\varepsilon}}\delta _{i,k}).  \label{a10}
\end{equation}
The quantities $\alpha _{\rm St}$, $\beta _{\rm St}$, and $\gamma _{\rm St}$
are simply
related to the reduced tensor amplitudes in Eq.~(\ref{a7a}); for instance,
\begin{equation}
\alpha _{\rm St}=\frac{-1}{\sqrt{3(2J_{a}+1)}}A^{0},\quad
\beta _{\rm St}^{q}=\frac{i%
}{\sqrt{2}}(-1)^{J_{b}-M_{b}}\left(
\begin{array}{ccc}
J_{b} & 1 & J_{a} \\
-M_{b} & q & M_{a}
\end{array}
\right) A^{1}.  \label{a11}
\end{equation}
{}From here on we will use the reduced amplitudes $A^{K}$ rather than
$\alpha _{\rm St}$, $\beta _{\rm St}$, and $\gamma _{\rm St}$, since the
irreducible tensor formalism makes it possible to write expressions in a
general form for all $K $.

At this stage let us assume that there is no nuclear spin and levels are
described by quantum numbers $J_{i},M_{i}$. Then we can replace the matrix
elements of $r_{q}$ in (\ref{a2}) in terms of reduced matrix elements, e.g.
\begin{equation}
\langle n|r_{q}|a\rangle =(-1)^{J_{n}-M_{n}}\langle n||r||a\rangle \left(
\begin{array}{ccc}
J_{n} & 1 & J_{a} \\
-M_{n} & q & M_{a}
\end{array}
\right) .  \label{a12}
\end{equation}

Using Eqs.~(\ref{a2}), (\ref{a7}), (\ref{a7a}), and (\ref{a12}), we obtain
an expression for $A^{K}$, with the sum (over $q,q^{\prime },$ and $M_{n}$)
of the product of the three 3j-symbols reduced to a 6j-symbol:
\begin{equation}
A^{K}=\sqrt{2K+1}\sum_{n}(-1)^{J_{a}+J_{n}}\left\{
\begin{array}{ccc}
J_{n} & J_{b} & 1 \\
K & 1 & J_{a}
\end{array}
\right\} \langle n||r||b\rangle \langle n||r||a\rangle \left( \frac{1}{%
E_{b}-E_{n}}+\frac{(-1)^{K}}{E_{a}-E_{n}}\right) .  \label{b2}
\end{equation}


\subsection{SI-E1 amplitude for nonzero nuclear spin}


We now include quantum numbers $F,M$ and $I$. First we can simply substitute
$J,M\rightarrow F,M$ in (\ref{b2}):
\begin{eqnarray}
A^{K} &=&\sqrt{2K+1}\sum_{n}(-1)^{F_{a}+F_{n}}\left\{
\begin{array}{ccc}
F_{n} & F_{b} & 1 \\
K & 1 & F_{a}
\end{array}
\right\} \times   \nonumber \\
&&\langle n,F_{n}||r||b,F_{b}\rangle \langle n,F_{n}||r||a,F_{a}\rangle
\left( \frac{1}{E_{b}-E_{n}}+\frac{(-1)^{K}}{E_{a}-E_{n}}\right) .
\end{eqnarray}
The dependence of the reduced matrix element on the quantum numbers $F$ is
described by the expression:
\begin{equation}
\langle n,F_{n}||r||a,F_{a}\rangle =(-1)^{J_{n}+I+F_{a}+1}\sqrt{%
(2F_{n}+1)(2F_{a}+1)}\left\{
\begin{array}{ccc}
J_{n} & F_{n} & I \\
F_{a} & J_{a} & 1
\end{array}
\right\} \langle n||r||a\rangle ,  \label{c2}
\end{equation}
so that
\begin{eqnarray}
A^{K} &=&(-1)^{2I+2F_{a}+F_{b}}\sqrt{\left( 2K+1\right) (2F_{n}+1)(2F_{a}+1)}%
\sum_{n}(-1)^{2J_{n}+F_{n}}(2F_{n}+1)\times   \nonumber \\
&&\left\{
\begin{array}{ccc}
F_{n} & F_{b} & 1 \\
K & 1 & F_{a}
\end{array}
\right\} \left\{
\begin{array}{ccc}
J_{n} & F_{n} & I \\
F_{b} & J_{b} & 1
\end{array}
\right\} \left\{
\begin{array}{ccc}
J_{n} & F_{n} & I \\
F_{a} & J_{a} & 1
\end{array}
\right\} \langle n||r||b\rangle \langle n||r||a\rangle \left( \frac{1}{%
E_{b}-E_{n}}+\frac{(-1)^{K}}{E_{a}-E_{n}}\right) .  \nonumber
\end{eqnarray}
Up to this point, our expressions have been perfectly general. From now on,
we introduce expressions that are specific to single valence-electron atoms.
In particular, we write the reduced matrix elements of $r$ explicitly in
terms of the single-electron angular momenta $l$ (orbital) and  $j$ (total):
\begin{equation}
\langle n||r||a\rangle =(-1)^{j_{a}+l_{an}-1/2}\sqrt{%
(2j_{n}+1)(2j_{a}+1)l_{an}}\left\{
\begin{array}{ccc}
l_{n} & j_{n} & \frac{1}{2} \\
j_{a} & l_{a} & 1
\end{array}
\right\} R_{n,a},  \label{b3}
\end{equation}
where $l_{an}\equiv \text{max}(l_{a},l_{n})$ and $R_{n,a}$ is the radial
integral. Combining these two equations gives
\begin{eqnarray}
A^{K} &=&-\sqrt{(2K+1)(2F_{a}+1)(2F_{b}+1)(2j_{a}+1)(2j_{b}+1)}%
\sum_{n}C_{F_{n},j_{n},l_{n}}^{K}D_{n}^{K},\medskip   \label{c3} \\
C_{F_{n},j_{n},l_{n}}^{K} &\equiv
&(-1)^{F_{n}+F_{b}+j_{a}+j_{b}}(2F_{n}+1)(2j_{n}+1)\times   \nonumber \\
&&\left\{
\begin{array}{ccc}
F_{n} & F_{b} & 1 \\
K & 1 & F_{a}
\end{array}
\right\} \left\{
\begin{array}{ccc}
j_{n} & F_{n} & I \\
F_{a} & j_{a} & 1
\end{array}
\right\} \left\{
\begin{array}{ccc}
j_{n} & F_{n} & I \\
F_{b} & j_{b} & 1
\end{array}
\right\} \left\{
\begin{array}{ccc}
l_{n} & j_{n} & \frac{1}{2} \\
j_{a} & l_{a} & 1
\end{array}
\right\} \left\{
\begin{array}{ccc}
l_{n} & j_{n} & \frac{1}{2} \\
j_{b} & l_{b} & 1
\end{array}
\right\} ,\medskip \medskip   \label{c4} \\
D_{n}^{K} &\equiv &(-1)^{l_{an}+l_{bn}}\sqrt{l_{an}l_{bn}}%
R_{n,a}R_{n,b}\left( \frac{1}{E_{b}-E_{n}}+\frac{(-1)^{K}}{E_{a}-E_{n}}%
\right) .  \label{c5}
\end{eqnarray}
Note that $D_{n}^{1}\propto E_{a}-E_{b}$; this is the suppression discussed
earlier, due to Bose statistics for photons.


\subsection{Summation over $F_{n}$ and $j_{n}$}


The general arguments for the suppression of the hyperfine SI-E1 amplitudes
suggest that there are cancellations in the sums over quantum numbers $F_{n}$
and $j_{n}$. To see this explicitly, note that in the nonrelativistic limit,
$D_{n}^{K}$ does not depend on $j_{n}$ or $F_{n}$. Application of the
Racah-Eliot relations\cite{Messiah} to the sums over $F_{n}$ and $j_{n}$
gives:
\begin{eqnarray}
\sum_{F_{n}}C_{F_{n},j_{n},l_{n}}^{K}
&=&(-1)^{F_{a}+I+K+j_{n}}(2j_{n}+1)\times  \nonumber \\
&&\left\{
\begin{array}{ccc}
K & j_{a} & j_{b} \\
I & F_{b} & F_{a}
\end{array}
\right\} \left\{
\begin{array}{ccc}
K & j_{a} & j_{b} \\
j_{n} & 1 & 1
\end{array}
\right\} \left\{
\begin{array}{ccc}
l_{n} & j_{n} & \frac{1}{2} \\
j_{a} & l_{a} & 1
\end{array}
\right\} \left\{
\begin{array}{ccc}
l_{n} & j_{n} & \frac{1}{2} \\
j_{b} & l_{b} & 1
\end{array}
\right\} , \medskip  \label{d1} \\
\sum_{j_{n},F_{n}}C_{F_{n},j_{n},l_{n}}^{K}
&=&(-1)^{F_{a}+I+j_{a}+j_{b}+l_{n}-1/2}\left\{
\begin{array}{ccc}
K & j_{a} & j_{b} \\
I & F_{b} & F_{a}
\end{array}
\right\} \left\{
\begin{array}{ccc}
K & l_{a} & l_{b} \\
\frac{1}{2} & j_{b} & j_{a}
\end{array}
\right\} \left\{
\begin{array}{ccc}
K & l_{a} & l_{b} \\
\ l_{n} & 1 & 1
\end{array}
\right\} .  \label{d2}
\end{eqnarray}
Note that both expressions turn to zero for $K>j_{a}+j_{b}$; this is
equivalent to the vanishing of a $\frac{1}{2}-\frac{1}{2}^{\prime }$ tensor
amplitude in the absence of nuclear spin. In addition expression (\ref{d2})
turns to zero for $K>l_{a}+l_{b}$. This is equivalent to the vanishing of $%
s-s^{\prime }$ vector (tensor) amplitudes in the absence of electron
(nuclear) spin.

Equations (\ref{d1}) and (\ref{d2}) make it possible to simplify the general
expression (\ref{c3}). If we neglect the dependence of $D_{n}^{K}$ on $F_{n}$
, we can use the sum (\ref{d1}). If we also neglect the dependence of $%
D_{n}^{K}$ on $j_{n}$, we can use the sum (\ref{d2}). This yields the
following expressions:
\begin{eqnarray}
A^{K} &=&-\sqrt{(2K+1)(2F_{a}+1)(2F_{b}+1)(2j_{a}+1)(2j_{b}+1)}\left\{
\begin{array}{ccc}
K & j_{a} & j_{b} \\
I & F_{b} & F_{a}
\end{array}
\right\} \times  \nonumber \\
&&\sum_{n}(-1)^{F_{a}+I+K+j_{n}}(2j_{n}+1)\left\{
\begin{array}{ccc}
K & j_{a} & j_{b} \\
j_{n} & 1 & 1
\end{array}
\right\} \left\{
\begin{array}{ccc}
l_{n} & j_{n} & \frac{1}{2} \\
j_{a} & l_{a} & 1
\end{array}
\right\} \left\{
\begin{array}{ccc}
l_{n} & j_{n} & \frac{1}{2} \\
j_{b} & l_{b} & 1
\end{array}
\right\} D_{n}^{K},\medskip \medskip  \label{d3} \\
A^{K} &=&-\sqrt{(2K+1)(2F_{a}+1)(2F_{b}+1)(2j_{a}+1)(2j_{b}+1)}\left\{
\begin{array}{ccc}
K & j_{a} & j_{b} \\
I & F_{b} & F_{a}
\end{array}
\right\} \left\{
\begin{array}{ccc}
K & l_{a} & l_{b} \\
\frac{1}{2} & j_{b} & j_{a}
\end{array}
\right\} \times  \nonumber \\
&&\sum_{n}(-1)^{F_{a}+I+j_{a}+j_{b}+l_{n}-1/2}\left\{
\begin{array}{ccc}
K & l_{a} & l_{b} \\
\ l_{n} & 1 & 1
\end{array}
\right\} D_{n}^{K},  \label{d4}
\end{eqnarray}
where $D_{n}^{K}$ is given by (\ref{c5}). The first of these equations is
correct up to the hyperfine corrections, while the second is correct only up
to the fine structure corrections.


\section{SI-E1 amplitudes for hyperfine transitions in alkali atoms}


We now apply these formulae to the $F_{a}=I-\frac{1}{2}\rightarrow F_{b}=I+
\frac{1}{2}$ transitions of the ground states $n_{0}s_{1/2}$ of alkali
atoms. As we have seen above, for such transition both vector and tensor
amplitudes are strongly suppressed. Here we want to obtain accurate
estimates of these amplitudes.

It is known that for all alkalis
\begin{equation}
|R_{n_{0}p,n_{0}s}|\gg |R_{np,n_{0}s}|,\quad n>n_{0}.  \label{e0}
\end{equation}
Thus, the dominant contribution to the sums (\ref{c3}), (\ref{d3}) and (\ref
{d4}) comes from the $p$ shell with the principle quantum number $n_0$.


\subsection{Vector amplitude}


For the case of interest where $l_{a}=l_{b}=0$, the nonrelativistic
expression (\ref{d4}) gives zero value for the amplitude $A^{1}$; to get a
correct value, we must use equation (\ref{d3}) which includes fine-structure
corrections. In order to avoid cancellation in the sum over $j_{n}$ we have
to take into account the dependence of the factor $D_{n}^{1}$ on $j_{n}$,
which can be written as:
\begin{equation}
D_{n}^{1}=\bar{D}_{n}^{1}+(j_{n}-l_{n})d_{n}^{1},  \label{e1}
\end{equation}
where $\bar{D}_{n}^{1}$ and $d_{n}^{1}$ do not depend on $j_{n}$. We have
seen that the contributions of $j_{n}=\frac{1}{2}$ and $j_{n}=\frac{3}{2}$
cancel each other for $\bar{D}_{n}^{1}$. Then obviously for $d_{n}^{1}$ they
should double each other, and we find:
\begin{equation}
A^{1}=\frac{4}{9}(-1)^{2I}\sqrt{3I(I+1)}\left\{
\begin{array}{ccc}
I-\frac{1}{2} & I+\frac{1}{2} & 1 \\
\frac{1}{2} & \frac{1}{2} & I
\end{array}
\right\} \sum_{np}d_{np}^{1}.  \label{e2}
\end{equation}
Note that the sum runs over states $np$ and does not include summation over $%
j_{n}$.

Now we have to find $d_{np}^{1}$. It follows from (\ref{e1}) that $%
d_{np}^{1}=D_{np_{3/2}}^{1}-D_{np_{1/2}}^{1}$ and it is not zero because of
the spin-orbit interaction $H_{\text{so}}$. There are diagonal and
off-diagonal contributions of $H_{\text{so}}$ to $d_{np}^{1}$:

\begin{itemize}
\item  The fine-structure splitting changes the energy denominators in (\ref
{c5}). If we define $\Delta _{\text{so},np}=E_{np_{3/2}}-E_{np_{1/2}}$ and
note that for the hyperfine transition $E_{b}-E_{a}=A_{n_{0}s}(I+\frac{1}{2})
$, where $A_{n_{0}s}$ is the hyperfine constant of the ground state, we get
for $d_{n}^{1}$:
\begin{equation}
d_{n}^{1}=-(2I+1)R_{np,n_{0}s}^{2}\frac{A_{n_{0}s}\Delta _{\text{so},np}}{%
(E_{n_{0}s}-E_{np})^{3}}.  \label{e3}
\end{equation}

\item  The spin-orbit interaction changes the radial integrals in (\ref{c5}):

\begin{equation}
\delta R_{np_{j},n_{0}s}=\sum_{n^{\prime }\ne n}\frac{\langle np|H_{\text{so}%
}|n^{\prime }p\rangle R_{n^{\prime }p,n_{0}s}}{E_{np}-E_{np^{\prime }}}.
\label{e4}
\end{equation}
\end{itemize}

Parametrically both mechanisms give the same smallness for $d_{n}^{1}$, but
numerically for alkali atoms the off-diagonal correction is suppressed.
Indeed, for the dominant term $n=n_{0}$, the correction to the radial
integral is numerically small because of the relation (\ref{e0}).

Thus the vector amplitude can be written:

\widetext
\begin{equation}
A^{1}=\frac{4}{9}(-1)^{2I+1}(2I+1)\sqrt{3I(I+1)}\left\{
\begin{array}{ccc}
I-\frac{1}{2} & I+\frac{1}{2} & 1 \\
\frac{1}{2} & \frac{1}{2} & I
\end{array}
\right\} R_{n_{0}p,n_{0}s}^{2}\frac{A_{n_{0}s}\Delta _{\text{so},np}}{
(E_{n_{0}s}-E_{n_{0}p})^{3}}.  \label{e5}
\end{equation}


\subsection{Tensor amplitude}


Here both expressions (\ref{d3}) and (\ref{d4}) turn to zero and we have to
return to the expressions (\ref{c3})--(\ref{c5}). Because of the relation (%
\ref{e0}) we again can restrict the summation to the shell $n_{0}p$. Then
the sum runs over $F=I\pm \frac{1}{2}$ and $j=\frac{1}{2},\frac{3}{2}$.
According to (\ref{d1}) the first sum vanishes because $j_{a}=j_{b}=\frac{1}{%
2}$, while (\ref{d2}) shows that the second sum vanishes because $%
l_{a}=l_{b}=0$. That means that all four coefficients $C_{F_{n},j_{n},0}^{2}$
differ only by their signs. So, it is sufficient to calculate, for example, $%
C_{I-\frac{1}{2},\frac{1}{2},0}^{2}$:

\begin{equation}
C_{I-\frac{1}{2},\frac{1}{2},0}^{2}=\frac{2}{3}(-1)^{2I+1}I\left\{
\begin{array}{ccc}
I-\frac{1}{2} & I+\frac{1}{2} & 1 \\
2 & 1 & I-\frac{1}{2}
\end{array}
\right\} \left\{
\begin{array}{ccc}
\frac{1}{2} & I-\frac{1}{2} & I \\
I-\frac{1}{2} & \frac{1}{2} & 1
\end{array}
\right\} \left\{
\begin{array}{ccc}
\frac{1}{2} & I-\frac{1}{2} & I \\
I+\frac{1}{2} & \frac{1}{2} & 1
\end{array}
\right\} .  \label{e6}
\end{equation}

It is clear that in order to avoid cancellations we have to include
hyperfine corrections to $D_{n}^{2}$ . Again, as in the case of vector
amplitude, we can neglect the correction to the radial integrals and
consider only corrections to the energy denominators. For the $n_{0}p_{3/2}$
state the quadrupole hyperfine structure should be included. That gives the
following result:
\begin{eqnarray}
A^{2} &=&\frac{8}{3}(-1)^{2I+1}I(2I+1)\sqrt{5I(I+1)}\left\{
\begin{array}{ccc}
I-\frac{1}{2} & I+\frac{1}{2} & 1 \\
2 & 1 & I-\frac{1}{2}
\end{array}
\right\} \left\{
\begin{array}{ccc}
\frac{1}{2} & I-\frac{1}{2} & I \\
I-\frac{1}{2} & \frac{1}{2} & 1
\end{array}
\right\} \times  \nonumber \\
&&\left\{
\begin{array}{ccc}
\frac{1}{2} & I-\frac{1}{2} & I \\
I+\frac{1}{2} & \frac{1}{2} & 1
\end{array}
\right\} R_{n_{0}p,n_{0}s}^{2}\left[ \frac{A_{n_{0}p_{1/2}}}{%
(E_{n_{0}s}-E_{n_{0}p_{1/2}})^{2}}+\frac{-A_{n_{0}p_{3/2}}+\frac{%
3B_{n_{0}p_{3/2}}}{(I(2I-1))}}{(E_{n_{0}s}-E_{n_{0}p_{3/2}})^{2}}\right] ,
\label{e7}
\end{eqnarray}
where $A_{n_{0}p_{j}}$ and $B_{n_{0}p_{3/2}}$ are the hyperfine constants
for the levels $n_{0}p_{j}$.


\subsection{Numerical results}


Equations (\ref{e5}) and (\ref{e7}) show that, in agreement with our general
arguments, the vector amplitude $A^{1}\sim \frac{\Delta _{{\rm hfs},n_{0}s}}{%
\Delta _{sp}}\frac{\Delta _{{\rm so},n_{0}p}}{\Delta _{sp}}a_{0}^{3}\propto
\frac{m_{e}}{m_{p}}Z^{3}\alpha ^{4}$, while the tensor amplitude $A^{2}\sim
\frac{\Delta _{{\rm hfs},n_{0}p}}{\Delta _{sp}}a_{0}^{3}$ $\propto \frac{%
m_{e}}{m_{p}}Z\alpha ^{2}$. This means that for light alkalis the tensor
amplitude dominates; however, the vector amplitude grows more rapidly with $%
Z $ and for heavy alkalis may even be larger than the tensor amplitude,
since the hyperfine constant of the ground $s$ state is much larger than the
hyperfine constants of the $p$ states.

To obtain numerical values of these amplitudes we need to know the radial
integrals $R_{n_{0}p,n_{0}s}$ and hyperfine constants $A_{n_{0}s}$ and $%
A_{n_{0}p_{j}}$. They are given in Table~\ref{tab1} together with our
results for the amplitudes $A^{1}$ and $A^{2}$. The uncertainty in these
values arises primarily from the two approximations we have made, namely:
neglecting intermediate $p$-states with $n>n_{0}$ in the sum of equation (%
\ref{c3}), and neglecting the $j$-dependence of radial matrix elements. We
estimate that these approximations introduce errors of $\lesssim 10\%$.

As stated in the introduction, the motivation for this calculation was the
effect of the SI-E1 amplitudes in experiments to measure PNC in hyperfine
transitions. In order to make the SI-E1 amplitudes comparable to the
parity-allowed M1 transitions between hyperfine sublevels (and thus to use
the SI-E1 amplitudes as the primary parity-allowed amplitude), it is
necessary that $A\sim \mu_B.$ Even for Fr, where $A$ attains its largest
value, this is achieved only for $E\sim 5\times 10^{9}$ V/cm! Such a
prospect is clearly unrealistic. On the other hand, it is also instructive
to compare the magnitude of the SI-E1 amplitudes to the PNC-induced E1
amplitudes. Using the values for the PNC amplitudes for hyperfine
transitions from Ref.\cite{Gorshkov} ($\kappa D_{Cs}\approx -3\times
10^{-13}ea_{0}$, $\kappa D_{K}\approx -1\times 10^{-14}ea_{0}$), we find
that in both cases the largest Stark-induced amplitude is comparable to the
PNC-induced amplitude only when the electric field $E\sim 3$ V/cm. Since
stray fields well below this level are easily controlled, systematic effects
due to uncontrolled SI-E1 amplitudes in experiments of the type proposed by
Gorshkov $et$ $al.$ should not be a serious problem, even for the lighter
alkalis.

We wish to thank D. Budker and S. Porsev for helpful discussions and useful
comments.


\mediumtext
\begin{table}[tbp]
\caption{Input data and numerical values for SI-E1 amplitudes. Unless
otherwise noted, data for $\Delta_{sp}$, $\Delta_{{\rm so},n_0 p}$, $A_{n_0
s}$, $A_{n_0 p_{1/2}}$, $B_{n_0 p_{3/2}}$, and $R_{n_0 s,n_0 p}$ are taken
from Ref.\protect\cite{Radzig&Smirnov}.}
\label{tab1}
\begin{tabular}{cccccccc}
&  & $^{7}$Li & $^{23}$Na & $^{39}$K & $^{85}$Rb & $^{133}$Cs & $^{221}$
Fr\smallskip \\ \hline
$I$ &  & $\frac{3}{2}$ & $\frac{3}{2}$ & $\frac{3}{2}$ & $\frac{5}{2}$ & $%
\frac{7}{2}$ & $\frac{5}{2}$\tablenotemark[1]\smallskip \\
$\Delta _{sp}$ & (cm$^{-1}$) & 14904 & 16965 & 13014 & 12698 & 11456 & 13081 %
\tablenotemark[2]\smallskip \\
$\Delta _{{\rm so},n_{0}p}$ & (cm$^{-1}$) & 0.34 & 17.2 & 57.9 & 238 & 554 &
1687 \tablenotemark[2]\smallskip \\
$A_{n_{0}s}$ & (GHz) & 0.402 & 0.886 & 0.231 & 1.01 & 2.30 & 6.21 %
\tablenotemark[1]\smallskip \\
$A_{n_{0}p_{1/2}}$ & (MHz) & 45.9 & 94.3 & 27.8 & 120.7 & 292 & 811 %
\tablenotemark[2]\smallskip \\
$A_{n_{0}p_{3/2}}$ & (MHz) & -3.06 & 18.7 & 6.1 & 25.0 & 50.3 & 66.5 %
\tablenotemark[2]\smallskip \\
$B_{n_{0}p_{3/2}}$ & (MHz) & -0.2 & 2.9 & 2.8 & 26.0 & -0.4 & -260 %
\tablenotemark[2]\smallskip \\
$R_{n_{0}s,n_{0}p}$ & ($a_{0}$) & 4.05 & 4.29 & 5.06 & 5.03 & 5.50 & 5.11 %
\tablenotemark[3]\smallskip \\
$A^{1}$ & (10$^{-6}a_{0}^{3}$) & 0.0085 & 0.72 & 2.0 & 70 & 940 &
2900\smallskip \\
$A^{2}$ & (10$^{-6}a_{0}^{3}$) & 20 & 27 & 20 & 185 & 1050 & 1350 \smallskip
\end{tabular}
\par
\tablenotemark[1]{}Ref. \cite{CocFr}
\par
\tablenotemark[2]{}Ref. \cite{Lu}
\par
\tablenotemark[3]{}Ref. \cite{Orozco}
\end{table}

\end{document}